\documentclass{article}
\usepackage{colortbl}
\newtheorem{example}{Example}%
\newtheorem{examplee}{Example}[section]
\newtheorem{definition}{Definition}%
\usepackage{algorithm2e}
\usepackage{amssymb}
\usepackage{multirow}
\usepackage{xspace}
\usepackage{epsfig}
\usepackage{subfig}
\usepackage{pifont}
\SetKwComment{Comment}{/* }{ */}
 \def\I{\mathcal{I}}

\def\MFIC{{\tt MFI-Compression}\xspace}
\def\MFI{{\tt MFI}\xspace}
\def\MFIs{{\tt MFI}\xspace}

\def\topk{{\tt TopK}\xspace}

\def\CFIs{{CFI}\xspace}
\def\STRC{{\tt STR-MFIC}\xspace}
\def\STRST{{\tt STR-Slice}\xspace}
\def\lcmmax{{\tt LCMmax}\xspace}
\def\smin{{\tt $S_{min}$}\xspace}

\def\T{\mathcal{T}}
\def\F{\mathcal{F}}

\def\CR{\mathcal{TD}}

\newcommand{\ie}{\textit{i}.\textit{e}.}
\newcommand{\lang}[1]{{\cal L}_{#1}}	
\newcommand{\theoryextension}[4]{\{#1 #2 #3\mid #4\}}

\usepackage{arxiv}

\usepackage[utf8]{inputenc} % allow utf-8 input
\usepackage[T1]{fontenc}    % use 8-bit T1 fonts
\usepackage{hyperref}       % hyperlinks
\usepackage{url}            % simple URL typesetting
\usepackage{booktabs}       % professional-quality tables
\usepackage{amsfonts}       % blackboard math symbols
\usepackage{nicefrac}       % compact symbols for 1/2, etc.
\usepackage{microtype}      % microtypography
\usepackage{lipsum}
\usepackage{graphicx}
\graphicspath{ {./images/} }

\title{An Efficient Heuristic Approach Combining
Maximal Itemsets and Area Measure for
Compressing Voluminous Table Constraints}

\author{
 Soufia Bennai \\
  LIMED Laboratory\\
  University of Bejaia\\
  Algeria\\
  \texttt{soufia.bennai@univ-bejaia.dz} \\
   \And
 Kamal Amroun \\
  LIMED Laboratory\\
  University of Bejaia\\
  Algeria\\
  \texttt{kamal.amroun@univ-bejaia.dz} \\
  \And
 Samir Loudni \\
  TASC (LS2N-CNRS)\\
  MT Atlantique\\
  France, Nantes R -44307 \\
  \texttt{samir.loudni@imt-atlantique.fr} \\
  \And
 Abdelkader Ouali \\
  GREYC\\
  niversity of Caen Normandie\\
  France \\
  \texttt{abdelkader.ouali@unicaen.fr} \\
}

\begin{document}
\maketitle
\begin{abstract}
Constraint Programming is a powerful paradigm to model and solve
com- binatorial problems. While there are many kinds of constraints,
the table constraint is perhaps the most significant-being the most well-
studied and has the ability to encode any other constraints defined
on finite variables. However, constraints can be very voluminous and
their size can grow exponentially with their arity. To reduce space
and the time complexity, researchers have focused on various forms of
compression. In this paper we propose a new approach based on maxi-
mal frequent itemsets technique and area measure for enumerating the
maximal frequent itemsets relevant for compressing table constraints.
Our experimental results show the effectiveness and efficiency of this
approach on compression and on solving compressed table constraints.
\end{abstract}

% keywords can be removed

\section{Introduction}
 
 Constraint Programming (CP) is a powerful paradigm to model and solve combinatorial problems. While there are many kinds of constraints, like automatas~\cite{Pesant04}, and MDDs (Multivalued Decision Diagrams)~\cite{b5}, the table constraint is perhaps the most significant being the most well studied and has the ability to encode any other constraints defined on finite domain (FD) variables. A canonical way of defining a FD constraint is simply to define the allowed (or disallowed) tuples of values, thus the constraint is defined as a table hence the term {\it table constraint}. 
Table constraints are widely used for modelling  applications of the real world, for instance, to encode user's preferences, to model database configuration problems, etc. Sometimes, table constraints provide the unique natural or practical way for a non-expert user to express her constraints. 

Over the last decade, research on table constraints has mainly focused on the development of fast algorithms to enforce generalized arc consistency (GAC), which is a property that corresponds to the maximum level of filtering when constraints are treated independently. There has been a large body of work on GAC algorithms for table constraints dating from the GAC4~\cite{MohrM88} and GAC-Schema~\cite{BessiereR97} (See \cite{YapXW20} for a detailed survey on techniques and algorithms for (G)AC on table constraints). Most of these algorithms propose different techniques for improving the implementation of the {\tt seekingSupports} function, it searches supports for each 
domain value. Among successful techniques, we find \cite{YapXW20} : 
\begin{itemize}
	\item {\it Residue Supports}. The idea of residue support is to record previously found supports, called residues, then seek supports from
	the residues to skip some checkings. Initially introduced for ensuring optimal complexity \cite{BessiereRYZ05}, newer GAC algorithms also use the residue idea, e.g. {STRbit}~\cite{WangXYL16} and {Compact-table} (CT)~\cite{VerhaegheLS17}. 
	
	\item Simple Tabular Reduction (STR)~\cite{Ullmann07} is one of
	the most successful techniques for filtering table constraints. 
	The idea of STR is to remove invalid tuples from tables as search 
	goes deeper, and restore them upon backtrack. 
	STR reduces the number of tuples of a table as search goes deeper, saving unnecessary tuple checks. Different variants of
	Simple Tabular Reduction (STR) have been proposed and proved to be quite competitive like STR2~\cite{Lecoutre11} and STR3~\cite{LecoutreLY15}.
	
	\item Bitwise Representation uses bit vectors to represent the domain and supports. It has been exploited more recently to the enforcement of GAC. Wang et al.~\cite{WangXYL16} propose a bitwise encoding together with the algorithm {STRbit}. {Compact-table} (CT) is another approach based on bitwise representation. Both approaches use bit vectors to record all valid tuples in a table (non-zero words in the
	bit vectors) during search. 
	
\end{itemize}

As said before, table constraints are important for modeling parts of many problems, but they admit practical boundaries because the memory space required to represent them may grow exponentially with their arity which can slow down their solving. To reduce space and the time complexity researchers have focused on various forms of compression. The intuition behind employing compact representations is that significant compression of tables should reduce running time for enforcing GAC. 
Multi-valued Decision Diagrams (MDDs)~\cite{b5} and bit-wise based algorithms are two examples of compact representations. More compact representations were also proposed to revise existing GAC algorithms, 
such as the c-tuples, short-supports, slice-tables, smart-tables~\cite{smart} and segmented-tables~\cite{segmented}. 
The corresponding GAC algorithms of different compact representations includes: {\tt GAC-ctuple}~\cite{b3}, {\tt STR2-C} and {\tt STR3-C} \cite{XiaY13}, and {\tt STRbit-C}~\cite{WangXYL16} for c-tuples; {\tt shortSTR2}~\cite{JeffersonN13} and {\tt shortCT}~\cite{VerhaegheLS17} for short-supports; {\tt STR-slice}~\cite{b1} for slice-table; {\tt smartSTR}~\cite{MairyDL15} and {\tt smartCT}~\cite{VerhaegheLDS17} for smart-tables. 

Other approaches propose to use data mining techniques for compressing table constraints, like the Microstructure Based Compression method \cite{b4}, 
{\tt sliced-table} \cite{b1} and {\tt FPTCM$^+$} \cite{a1}. The {\tt sliced-table} approach exploits an FP-Tree structure to enumerate the frequent itemsets from a table constraint and uses the notion of the savings that can be offered by an itemset to select the frequent itemsets that are relevant for compression. The {\tt FPTCM$^+$} approach is an improvement of the {\tt sliced-table} method, it uses the concept of compression rate to enumerate frequent itemsets that are more relevant for compression.

In this paper, we go one step further in exploiting data mining approaches to compress table constraints. We propose to use the maximum frequent itemset (MFI) to cover a maximum number of variables in the scope of the table constraint. This allows to reduce the size of the tuples in the resulting
compressed tables. To achieve better compression, we select the \MFI covering a maximum number of tuples (\ie high frequency). However, the larger the \MFI, the lower the frequencies. A better compromise between the length and the frequency of \MFI is to exploit the area measure (the product of the length of an itemsets and its frequency value) such that we select the \MFI with higher area values. To mine the set of \MFI, the value of the minimum frequency threshold \smin has to be fixed, for this we dynamically fix for each table constraint the value of \smin  by using the \topk approach. Finally, the relevance and the effectiveness of our approach is highlighted through a set of experiments on benchmarks downloaded from \emph{https://bitbucket.org/pschaus/xp-table/src/master/instances/}. The obtained results are very promising.
The remainder  of this paper is organized as follows. In Section~\ref{prel}, we give some definitions 
related to Constraint Satisfaction Problems (CSPs) and frequent itemsets mining. Section~\ref{rel} reviews some related works. 
Section~\ref{contrib} is devoted to our proposition called \MFIC. In section~\ref{complx} we calculated the time complexity of our approach. Section ~\ref{solv} explains how solving the compressed constraints. Experiments, carried out in this work, are presented in Section~\ref{exp}. We conclude with some remarks and avenue for future works in Section~\ref{conc}. 

\section{Background}
In this section some concepts related to Constraint Satisfaction Problems (CSPs) and Data Mining are formally defined~\cite{guns,han,maamar}.

\subsection{Constraint Satisfaction Problem}
Constraint Satisfaction Problem (CSP) was formally defined by U. Montanari ~\cite{b2} as  a finite set of variables $X$=  $\{x_{1},\cdots,  x_{n}\}$ with finite
domains $\mathcal{D} = \{D_1, \ldots, D_n\}$. Each $D_i$ is the set of possible values that can be assigned to $x_i$, 
and a finite set of constraints $\mathcal{C}$=  $\{c_{1},\cdots,  c_{m}\}$. A constraint $c_i \in \mathcal{C}$ is a pair $(S(c_{i}), R(c_{i}))$, 
where:
\begin{itemize}
    \item $S(c_{i})\subseteq X$ is the scope of the constraint $c_i$. It represents the set of variables involved in $c_{i}$ ;
    \item $R(c_{i}) \subseteq \prod_{x_{k}\in S(c_{i})} D _{k}$ is a relation that defines the set of tuples allowed for the variables of $c_{i}$.
\end{itemize}

The size of the set $S(c_{i})$ is called the {\it arity} of the constraint $c_i$. A unary constraint is a constraint of arity one, a binary constraint is a constraint of arity two, a non-binary constraint is a constraint of arity greater than two. 

The size of a constraint relation $R(c_{i})$ is the product of the arity of $c_{i}$ by the number of tuples in $R(c_{i})$. The relation of a constraint can be specified {\it extensionally} by explicitly listing its acceptable tuples, or {\it intensionally} by specifying an expression that tuples in the constraint must satisfy.  Example~\ref{csp} shows a  CSP instance  defined in extension. \\
%\begin{example}
\begin{example}
\label{csp}
Consider the following CSP defined in extension:\\
$X = \{x_{0}, \ldots, x_{4}\}$, 
$D = \{D_{0}, \ldots, D_{4}\}$ where, 
$D_{0} = \{0, 1\}$,  $D_{1} = \{0, 1, 2\}$,  $D_{2} = \{0, 1, 2\}$, $D_{3} = \{0, 1, 2, 3\}$, 
$D_{4} = D_{2}$. \\ 
$C = \{c_{0}\}$ where $c_{0} = ((x_{0}, x_{1}, x_{2}, x_{3}, x_{4}), R(c_{0}))$ and \\
$R(c_{0})$ $=$ $\{(0$ $0$ $0$ $0$ $2),$ $(0$ $0$ $0$ $1$ $2),$ $(0$ $2$ $0$ $2$ $0),$ $(0$ $0$ $1$ $1$ $2),$ $(0$ $0$ $1$ $2$ $0),$ $(0$ $0$ $1$ $3$ $2),$ $(1$ $0$ $2$ $1$ $1),$ $(1$ $0$ $2$ $3$ $0),$ $(1$ $1$ $2$ $0$ $1),$ $(1$ $1$ $2$ $2$ $2),$ $(1$ $1$ $2$ $3$ $0\}$. 
%$X = \{x_{0}, \ldots, x_{5}\}$, 
%$D = \{D_{0}, \ldots, D_{5}\}$ where, 
%$D_{0} = \{0, 1, 2, 3, 4, 5, 6\}$,  $D_{1} = \{1, 2, 3, 4\}$,  $D_{2} = \{0, 1\}$, $D_{3} = \{0, 1, 2\}$, 
%$D_{4} = D_{5} = D_{2}$. \\ 
%$C = \{c_{0}\}$ where $c_{0} = ((x_{0}, x_{1}, x_{2}, x_{3}, x_{4}, x_{5}), R(c_{0}))$ and \\
%$R(c_{0})$ $=$ $\{(0$ $1$ $0$ $0$ $0$ $0),$ $(0$ $1$ $0$ $1$ $0$ $0),$ $(0$ $1$ $0$ $2$ $0$ $0),$ $(0$ $2$ $0$ $0$ $0$ $0),$ $(0$ $2$ $1$ $0$ $0$ $0),$ $(0$ $2$ $0$ $1$ $0$ $0),$ $(0$ $2$ $0$ $2$ $0$ $0),$ $(0$ $2$ $0$ $2$ $0$ $1$)$\}$. 
\end{example}

An {\it assignment} is a pair $(x_i, a)$, which means that the variable $x_i \in X$ is assigned the value $a \in D_i$. A {\it partial} assignment (noted $\overrightarrow{A_{i}}$) is a set of assignments to distinct variables in $X$. A {\it complete} assignment is an  assignment to all variables in $X$.
We say that a partial assignment satisfies a constraint $c_i$ if the restriction of the assignment 
to the scope $S(c_{i})$ is an acceptable (satisfying) tuple. A solution to a CSP instance $P = \langle X, D, C\rangle$ is a complete assignment that {\it satisfies} all constraints of $C$. 
Solving a CSP $P$ consists in checking whether $P$ admits at least one solution. It is a NP-hard problem.
If no solution exists, the CSP is said to be inconsistent or unsatisfied.

There exists many complete and incomplete techniques for solving CSPs. Most “efficient” complete methods rely on a depth-first search with backtracking combined with Constraint propagation and variable/value ordering heuristics. In the worst case, their time complexity is in $O(d^{n})$ 
(with $n$ is the number of variables and $d$ is the size of the largest domain) 
while being generally linear in space. 

Depth-First Search methods explore a search tree in a systematic way by recursively choosing the next unassigned variable to assign and by choosing a value in its domain for the assignment (the branch part) until a solution is found or it can be proved that the subtree rooted at the current search
node has no solution. At each search node, constraint propagation is performed to filter the domains of variables so that values that cannot be part of a solution are removed from the domains of unassigned variables. When one domain of a variable becomes empty, this means that the lastly instantiated variable conducts some constraints to be violated. Hence, the algorithm needs to backtrack in order to consider another possible value for this variable. Most solvers maintain generalized arc consistency for the table constraint.

\begin{definition}[Support] 
A support of a constraint $c \in C$ is a set of assignments
to exactly the variables in $S(c)$ such that $c$ is satisfied. A support of $c$ that
includes the assignment $(x_i, a)$ is called a support of $x_i$ in $c$. 
\end{definition}

\begin{definition}[Generalized arc consistency (GAC)] 
	A constraint $c$ is GAC if there exists a support for all values in the current domains of the variables in $S(c)$. A CSP is GAC if all of its constraints are GAC.
\end{definition}

\subsection{Frequent itemset mining}
Let $\I$ be a set of $n$ distinct literals called items, an itemset (or \emph{pattern}) is a non-null subset of $\I$. 
The language of itemsets corresponds to  $\lang{\I}$ = $2^{\I}$ $\backslash$ $\emptyset$. 
A \emph{transaction data set} is a multi-set of $m$ itemsets of $\lang{\I}$. 
Each itemset, usually called a \emph{transaction} or object, is a data set entry.  

Let $\CR$ %$\T$ 
be  a transaction data set, $u \in \lang{\I}$ be an itemset, and $match\footnote{For an itemset $u \in \lang{\I}$ and a transaction $t$, $match(p,t) = true$ iff $p$ covers the transaction $t$.}: \lang{\I} \times \lang{\I} \mapsto \{true,false\}$ a matching operator. Table~\ref{tt0} presents an example of a transaction data set $\CR$ where each tuple (transaction) $t_{i}$ is described by items denoted $A, \cdots, E$.\\

\begin{table}[t] \centering
\begin{center} 	
\caption{Transactional dataset $\CR$.}\label{tt0}
	\begin{tabular}{llllll}
	\toprule
		tid & \\%\multicolumn{5}{l}{items}\\
		\midrule
		$t_{0}$& C & D& E& A& B\\
		$t_{1}$&E & B& C& D& \\
		$t_{2}$&E & C & D& & \\
		$t_{3}$&D & A&  C& E & \\
		$t_{4}$&E & C& A& B& \\
	\bottomrule
	\end{tabular}
\end{center}
\end{table}

\begin{definition}[Coverage and Frequency]
\label{def:cov:freq} 
Let $\CR$ be a transaction database over a set of items $\I$, the set of identifiers of tuples in which an itemset $u$ appears is called the coverage of $u$: 
\begin{equation}
    cover(u) = \{t \in \T  \vert \forall i \in u, (i,t) \in \CR \}
\end{equation}
The frequency of an itemset $u$ is the size of its coverage: 
$freq(u)$ = $\mid cover(u) \mid$.
\end{definition}
 \begin{example}
\label{def:ex1}
Consider the transaction data set $\CR$ in Table~\ref{tt0}. 
We have for $u = EC$, $cover(u) = \{t_{0}, t_{1}, t_{2}, t_{3}, t_{4}\}$ and $freq(u) = 5$. 
\end{example}
\begin{example}
	\label{exmpl1}
By considering the transaction data set $\CR$ in Table~\ref{tt0} and $S_{min} = 2$, the itemset $u = EC$ is a frequent itemset because $freq(u) > S_{min}$. 
\end{example}

\begin{definition}
Let $\CR$ be a transaction database over a set of items $\I$,
and let $S_{min}$ be a minimal support threshold. We note the collection of frequent itemsets in $\CR$ 
with respect to $S_{min}$ by: $\mathcal{F}(\CR,S_{min}) = \{u \in \lang{\I} \mid freq(u) \ge S_{min}\},$ 
or simply $\mathcal{F}$ if $\CR$ and $S_{min}$ are clear from the context. \\
\end{definition}

\begin{definition}[Frequent Itemset Mining Problem]
Let $S_{min}$ be a minimal support threshold. The frequent itemset mining problem is the computation of the set of all itemsets $u$ having frequency in the data set exceeding $S_{min}$ : $freq(u) \geq S_{min}$. 
\end{definition}

When a database is very dense or the value of the minimal support $S_{min}$ is set too low, mining all the frequent itemsets can be impractical because of the huge number of possible frequent itemsets. 
To limit the number of output, several reduction techniques based condensed representations of patterns have been proposed in the context of the frequency measure~\cite{BoulicautBR03,Bayardo98,MannilaT97,PasquierBTL99}. The most popular ones are {\it closed} and {\it maximal} itemsets. 

\begin{definition}[Closed frequent itemset]
\label{clos}
	A frequent itemset $u_i \in \F(\CR, S_{min})$ is closed iff
$\forall \, u_j \in \lang{\I}, \, \, u_j \subsetneq u_i \Rightarrow freq(u_j) < freq(u_i)$.   
\end{definition}

\begin{example}
%\textbf{Example~4} 
\label{exmpl2}
Consider $S_{min}$ $=$ $2$. From Table~\ref{tt0}, we get four frequent closed itemsets which are: $CE \langle 5 \rangle$, $CDE \langle 4 \rangle$, $ACE \langle 3 \rangle$, $AD \langle 2 \rangle$, $BCE \langle 3 \rangle$, $BCDE \langle 2 \rangle$, $ABCE \langle 2 \rangle$. The value between $ \langle \rangle$ indicates the frequency of an itemset.
\end{example}

Since the collection of all frequent itemsets is {\it downward closed}, meaning that any subset of a frequent
itemset is frequent, it can be represented by its maximal elements, the so called {\it maximal frequent itemsets}.

\begin{definition}[Maximal frequent itemset]
\label{def-maxitemset}
A frequent itemset $u_i \in \F(\CR, S_{min})$ is called maximal iff 
$\forall \, u_j \in \lang{\I}, \, \, u_j \supsetneq u_i
\Rightarrow freq(u_j) < S_{min}$. 
\end{definition}
\begin{example}
%\textbf{Example~5} 
\label{exmpl3}
In Table~\ref{tt0}, if we impose that $S_{min} = 2$, we obtain two maximal frequent itemsets: $BCDE \langle 2 \rangle$, $ABCE \langle 2 \rangle$ and $ACDE \langle 2 \rangle$.
\end{example}

Other studies attempt to integrate user preferences into the mining task in order to limit the number of extracted patterns such as the \topk pattern mining approaches~\cite{KeCY09,WangHLT05}. 
By associating each pattern with a rank score, such as frequency, this approach returns an ordered list of the $k$ patterns with the highest score to the user.

\begin{definition}[\topk frequent itemsets]
\label{topk}
Let $k$ be an integer. \topk  w.r.t. the frequency measure is the set of $k$ best frequent itemsets:
\begin{equation}
\begin{array}{ll}
\theoryextension{u}{\in}{\l
 %\textbf{Example~3} 
ang{\I}}{ & freq(u)\ge S_{min}\, \, \land\ 
 \not \exists x_1,\ldots,x_k\in\lang{\I}:\ \forall 1 \le j \le k,freq(x_j) > freq(u) }
\end{array}
\end{equation}
\end{definition}
\begin{example}
%\textbf{Example~6} 
	\label{exmpl4}
	In our running example (with $S_{min} = 4$), the $top$-7 frequent itemsets are: $E \langle 5 \rangle, C \langle 5 \rangle$, $D \langle 4 \rangle$, $EC \langle 5 \rangle, ED \langle 4 \rangle, CD \langle 4 \rangle, ECD \langle 4 \rangle$. 
\end{example}

Regarding the algorithmic approaches for mining closed itemsets, much effort on developing sophisticated algorithms have been expended. {\tt LCM} (Linear time Closed frequent itemset Mining)~\cite{lcm} is one of the most prominent and performer algorithm for this task. {\tt LCMmax}~\cite{lcm} is an extension of {\tt LCM} dedicated to mine maximal frequent itemsets (\MFI). Its main feature is to have a linear complexity w.r.t the number of closed itemsets. \lcmmax enumerates the set of all closed frequent itemsets (\CFIs) by backtracking and exploits pruning and maximality checking techniques to accelerate the computation time and to avoid storing the \MFIs previously found in memory. 

\medskip
In addition to the frequency, other interestingness measures, like the {\it area}, can be exploited. 

\begin{definition}[area of itemset]
\label{area}
The area of an itemset $u$ is the size of the itemset $\vert u\vert$ multiplied by its frequency $freq(u)$ : 
\begin{equation}
    area(u) = \vert u \vert * freq(u)
\end{equation}
\end{definition}

In the sequel, we show how to take advantage of maximal patterns to compress table constraints by selecting those that maximize the area measure. 

\subsection{Constraint based compression by itemset mining}
In this section we show how a table constraint $R(c_{i})$ associated with  a constraint $c_{i}$ can be represented as a transactional dataset $\CR_{c_i}$. Then, we show how to compress $R(c_{i})$ using 
itemset mining techniques.

Let $P=\langle X, D, C\rangle$ be a CSP and $R(c_{i})$ be a table constraint associated with 
a constraint $c_{i} \in C$. The transactional dataset $\CR_{c_i}$ is defined as
follows: 
\begin{itemize}
    %\item tuples sorted according to decreasing frequencies
    \item [(i)] the union of the domains of the variables in the scope of $c_{i}$ represents the set of items of $\I$, 
    \item [(ii)] the set of values involved in the tuple $t\in R(c_{i})$ forms a transaction in $\T$.
\end{itemize}

In this context, an itemset represents an assignment of some variables involved in the scope of $c_{i}$. Table~\ref{tdc} shows the transactional dataset  $\CR_{c_0}$ associated with the table constraint  $R(c_{0})$ of Example~\ref{csp}. If we consider $S_{min} =2$, the following assignements: $\langle x_{0}=1, x_{2}=2\rangle$ represent an example of a frequent itemset of Table~\ref{tdc} that covers the tuples 
$t_7, t_8, t_9$ and $t_{10}$.\\ 
%from  Table~\ref{tab:itemsets:exple} reports the maximal frequent itemsets extracted from  $\CR_{c_0}$. 
%%%%%%%%%%%%%%%%%%%%%%%%%%%%%%%%%%%%%%%%%%%%%%%%%%%%%%%%%%%%%%%%%%%%%%
% taux compression : 1-(8*5+9)/55= 0.109 ~ 10%
\begin{table}[t]
\caption{Running example.}	\label{tdc}
\begin{center}
	\begin{minipage}{.5\textwidth}
	\subfloat[Transactional dataset $\CR_{c_0}$\label{td} ]{
	\begin{tabular}{llllllll}
		\toprule
		TID &$x_{0}$&$x_{1}$&$x_{2}$&$x_{3}$&$x_{4}$\\
		\midrule
		$t_{0}$&0&0&0&0&2\\
		$t_{1}$&0&0&0&1&0\\
		$t_{2}$&0&2&0&2&0\\
		$t_{3}$&0&0&1&1&2\\
		$t_{4}$&0&0&1&2&0\\
		$t_{5}$&0&0&1&3&2\\
		$t_{6}$&1&0&2&1&1\\
		$t_{7}$&1&0&2&3&0\\
		$t_{8}$&1&1&2&0&1\\
		$t_{9}$&1&1&2&2&2\\
		$t_{10}$&1&1&2&3&0\\
		\bottomrule
		\end{tabular} }
%	\footnotetext{Transactional dataset $\CR_{c_0}$}\label{tdc} 
\end{minipage}
\def\arraystretch{0.8}
	{
	\subfloat[An entry of $R(c_0)$.\label{tdc:entry}]{
	    \begin{tabular}{p{2.5cm}p{1.5cm}}
		  \toprule
			$x_{0}$ $x_{1}$ $x_{2}$& \multicolumn{1}{l}{$x_{3}$ $x_{4}$} \\
		  \midrule
			& 1$\:$ 1  \\
	0 $\:$ 0 $\:$ 1  & 2$\: $  0  \\
			& 3$\: $  2 \\
          \bottomrule
         % \footnotetext{An entry of $R(c_0)$.}
	\end{tabular}}
	
}
\end{center}
\end{table}

The main idea behind the use of pattern mining to derive a compact representation of the table constraint is to use frequent itemsets extracted from the transaction dataset as a summary of a set of transactions. These transactions are replaced by each frequent itemset that covers them. 
The resulting compressed constraint relation consists  of  a  set  of  entries  where  each  entry
contains   an   itemset   and   its   corresponding   sub-table.
\medskip
\begin{definition}[Sub-table]
	The sub-table $St$ associated with an itemset $u$ of a constraint $c_{i}$ consists in the remaining parts of tuples in the coverage of $u$ after removing $u$ from each tuple.
\end{definition}
%\medskip
\noindent
\begin{definition}[Entry]
\label{entry}
	An entry for a table constraint $R(c)$ is a pair $(u, St)$ such that $u$ is a frequent itemset and $St$ its corresponding sub-table. 
\end{definition}
%%

 %\textbf{Example~8 } 
 \begin{example}
	Table~\ref{tdc}(b) shows the entry corresponding to the itemset $u = \{x_{0} = 0, x_{1} = 0, x_{2} = 1$ and its resulting sub-table.
\end{example}

\begin{definition}[Default table]
A default-table for a table constraint $R(c)$ is a table that contains all tuples that can not be compressed with the mined frequent itemsets.
\end{definition}

%\textcolor{red}{TODO: the concept of default table disappeared, while it is used in Table 4}

%\textcolor{red}{TODO: the definition of the compressed table is not correct as we can also include the default table}
\begin{definition}[Compressed table constraint]
\label{cmc}
A compressed table constraint $R^*(c)$ of $R(c)$ is represented by a set of entries associated to the set of non-overlapping frequent itemsets and a default-table.

\end{definition}
Let $u$ be a frequent itemset, $f$ its frequency and $T$ the set of compressed transactions. Let $size_a$ (resp. $size_b$) be the size of $T$ after (resp. before) compression. To assess the quality of a summary $u$ of a set of transactions $T$, we define the following metric: %\textcolor{red}{TODO: The compression ratio is not defined}
\noindent
The size $size_{a}$ of the transactions after their compression is equal to the length of $u$ plus the size of its corresponding sub-table. The size of the sub-table is obtained by multiplying the arity of the sub-table ($arity - |u|$) by the frequency of $u$ : $size_{a} = |u| + (arity - |u|) * f$.
The compression ratio of a $T$ w.r.t. itemset $u$ is obtained as follows: $Rate$ = $1- \frac{size_{a} }{size_{b} }$ where $size_{b} = arity * f$.

\section{Related Works}
In this section, we review some compact representations proposed in the literature for table constraints. 

\medskip
\noindent
\textbf{Katsirelos and Walsh}~\cite{b3} have proposed first a compact representation of constraint relations. They exploit a decision tree to represent the original constraint relation as a disjunction of tuples. Then, they extract from this decision tree a set of compact tuples called c-tuples that will be used to represent the constraint relation as a conjunction of c-tuples. Thus a compact representation can exponentially reduce the size of a constraint relation and the time complexity required to enforce GAC (Generalized Arc Consistency). 

%proposed a first approach for compressing large arity table constraints using decision trees. A table constraints which is a disjunction of tuples is represented by a set of conjunction of new compressed tuples (c-tuples), leading to a more compact representation such that each c-tuple can represent $n$ no compressed tuples. 
\medskip
\noindent
\textbf{Cheng et al}. \cite{b5} have proposed a new form of compression based on Multi-valued Decision Diagrams (MDD). The size of a tree is often smaller than the size of the constraint relation. That is why the authors have proposed such a structure to perform an effective support checking. They also proposed to merge the identical sub-tries in the decision tree to reduce the time required for support checking, thus obtaining a directed acyclic graph (DAG), called a multi-valued decision diagram (MDD). 
Two notable algorithms using MDDs as main data structure are {\tt mddc} \cite{b5} and {\tt MDD4R} \cite{PerezR14}. The former does not modify the decision diagram and performs a depth-first search of the MDD during propagation to detect which parts of the MDD are consistent or not. {\tt MDD4R} dynamically maintains the MDD by deleting nodes and edges that do not belong to a solution.

%The resulting MDD is then recursively traversed in order to seek for each assignment.

%\textbf{Mairy et al}. ~\cite{smart} proposed another form of compression of constraint relations called smart tables. It represents a constraint relation as a set of smart tuples. Smart tuples contain simple arithmetic constraints.\\
\medskip
\noindent
\textbf{Jabbour et al}. \cite{b4} have proposed a SAT based approach for compressing table constraints. They introduced two new rewriting rules for reducing the size of the constraint network as well as the size of the constraint relations while preserving the original structure of the constraints. They used closed itemsets to compute a summary of tuples of each table constraint.

\medskip
\noindent
\textbf{Some variants of STR algorithms} work on compressed
table representations. {\tt STR2-C} and {\tt STR3-C} \cite{XiaY13} 
works on the Cartesian Product representation (c-tuple) of tuples to compress tables. 

\medskip
\noindent
\textbf{Wang et al.}~\cite{WangXYL16} proposes a bitwise encoding of 
the dual table representations together with the algorithms {\tt STRbit} and {\tt STRbit-C}. To get the bitwise representation, the original table is first partitioned so that each subtable have w tuples where $w$ corresponds to the natural word size of processor with $O(1)$ bit vector operations. Compact-table (CT) is another state-of-the-art algorithm,
also based on bitwise representation. Both CT and 
STRbit(-c) use bit vectors to record all valid tuples in a table 
(non-zero words in the bit vectors) during search. 

\medskip
\noindent
\textbf{Gharbi et al}.~\cite{b1} have introduced sliced-table \cite{b1}, 
a new compression method based on FP-Tree structure to enumerate the frequent itemsets relevent for compressing constraint relations. The proposed approach takes as input a constraint relation to compress and returns a set of entries and a default table which contains tuples that are not compressed. To decide either an itemset $u$ of the FP-Tree corresponding to a constraint relation is relevant for compression or not, the authors proposed to compute the savings that can be obtained by factoring $u$. The saving of $u$ is computed by the following formula : $|u|*(freq(u)-1)$, where $u$ is an itemset and $freq(u)$ its frequency.  

\medskip
\noindent
 \textbf{Audemard et al}.~\cite{segmented} introduced the notion of segmented table that generalize compressed tables. A segmented constraint is represented with a set of segmented tuples. Where each segment of a segmented table constraint can be  represented with universal values (*), ordinary values or sub-tables. Then authors proposed an algorithm for  enforcing GAC on segmented tables.

 \section{A new heuristic approach based on maximal patterns for compressing table constraints}
  % SUGGESTION :  schéma global
% décomposer la méthode en étape, chaque étape est représentée par un algorithme
%

In this section, we detail our heuristic approach, called \MFIC, based on maximal itemsets for compressing table constraints.

\subsection{Our approach in nutshell}
%Iterating over compressed table constraints during the search can accelerate the filtering process by avoiding redundancy in the set of tuples of each table constraint. 
%Frequent itemsets constitute good candidates to reduce the entries of table constraints. 
To achieve a better compression, our approach first selects {\bf maximal itemsets} (\MFI). Indeed, considering itemsets involving more variables in the scope of the table constraint allows to reduce the size of the tuples in the resulting compressed tables. Moreover, we aim to select those covering a large number of tuples in the table, that is, those with high frequency. 
However, the larger the \MFI, the lower the frequencies. It is thus necessary to ensure a better compromise between these two criteria, i.e. length and frequency.  
To that end, we propose to exploit the area measure to achieve such a compromise. %Indeed, maximizing both the length and the frequency of an \MFI is equivalent to maximize its area. 
Consider, for instance, the set of \MFI in Table~\ref{ex:tab:max} extracted from our running example in Table~\ref{td}. The first \MFI with size three covers $3$ tuples, while the second \MFI with size four covers only $2$ tuples. 
%\textcolor{red}{TODO: report the compression value for both patterns. }

According to the area measure, it would be more interesting to select the first \MFI because its area (equal to $9$) and its compression ratio (equal to $10.9\%$) are larger than the second \MFI's area (equal to $8$) and compression ratio (equal to $7.6\%$), respectively. 

\begin{algorithm}[hbt!]
\caption{\MFIC}\label{cont}
\KwData{ $\CR$: table constraint $c$ to compress, $k$: number of \CFIs to fix $S_{min}$.}
\KwResult { $R^c$: compressed table constraint.}
$\mathcal{S} \leftarrow  {\tt Select} (\mathcal{M})$ \Comment*[r]{see section \ref{i}}
$\mathcal{F}^k \leftarrow \topk(\CR, k)$ \Comment*[r]{see section \ref{i}}
$S_{min} \leftarrow \min_{F \in \mathcal{F}^k} freq(F)$; \\%\Comment{The minimum frequency threshold $S_{min}$}
$\mathcal{M} \leftarrow {\tt LCMmax(\CR,S_{min})}$ ; \\ %\Comment{see section \ref{ii}}
create the compressed table constraint $R^c$;\\ %\leftarrow {\tt CompressedTable}(\mathcal{S},\CR)$ \Comment{see section \ref{iv}}
Return $R^c$
\end{algorithm}

Consider again the \MFI in Table~\ref{ex:tab:max}. We can see that the tuple $t_{10}$ appears in the coverage of the first and the sixth \MFI, while it can be compressed using only one \MFI. To prevent compressing tuples more than once, we remove all the compressed tuples from the coverage of the remaining non yet selected \MFI. 
The main steps of our \MFIC heuristic are depicted in Algorithm~\ref{cont}:
\begin{enumerate}
    \item[(a)] select the best value for the minimum frequency threshold $S_{min}$; 
    \item [(b)] extract the \MFI using \lcmmax with the value of $S_{min}$ found in step (i); 
    \item [(c)] choose heuristically the most relevant \MFI maximizing the area measure; 
    \item [(d)] create the compressed table constraint $R^c$. 
\end{enumerate}

\subsection{Setting the parameter $S_{min}$ and mining candidate \MFI}
\label{i}
Finding the suitable minimum threshold value \smin for each table constraint is challenging. If its value is maintained too low, too many \MFI can be mined, and the relevant ones can hardly be found among the resulting massive set of \MFI. Similarly, if the value of \smin is too high, too few number of \MFI can be generated, and some \MFI relevant for compressing the table constraint can be missed. To generate a good set of \MFI candidates regardless of the table constraints, \MFIC dynamically fix for each table constraint the value of \smin by using the \topk approach~(line $1$, Algorithm~\ref{cont}). Let $k$ be a user-defined value. We first generate the top$-k$ closed most frequent itemsets, then we set \smin to the lower frequency value among all the mined \CFIs (line $2$,  Algorithm~\ref{cont}). Finally, using the {\tt LCMmax}~\cite{lcm} method, we extract from the constraint table all the \MFI   w.r.t. \smin (line $3$, Algorithm~\ref{cont}). 
For instance, If we consider Table~\ref{tdc} and $k=10$, according to our \topk approach, \smin = 2.

\SetKwRepeat{Struct}{struct \{}{\}}%
\SetKwRepeat{Function}{function \{}{\}}%
\begin{algorithm}[hbt!]
\caption{{\tt Select()}}\label{filter}*
\KwData{$\mathcal{M}$: List of \MFIs.}
\KwResult{ $\mathcal{S}$: List of \MFIs selected for compression.}

\Struct{E}{
  $p$ : \textsc{Itemset};\\
  $area$ : \textsc{Float};\\
  \Function{Comparator}{E other}{
    \If{$area > other.area$}{
        {\bf Return} true;\\
    }
    {\bf Return} false;
  }
}

$L \leftarrow$ new List of E;
\For{$u \in \mathcal{M}$}{
    $area \leftarrow size(u) * cover(u) $;\\
    $L.insert(new~E(u, area))$;\\
}
$L.sort()$ \Comment*[r]{{\scriptsize\tt elements of $L$ are set in a decreasing order based on the attribute area.}}
\While{$L \neq \emptyset$}{
    u $\leftarrow L.first()$;\\
    $\mathcal{S}.add(u)$;\\
    $L.remove(u)$ \Comment{{\scriptsize\tt remove the selected itemset $u$ from the list $L$.}}
     $i \leftarrow 0$;\\
    \While{$i < L.size()$}{
        \eIf{$cover(u) \cap cover(L.get(i).p)$}{
          $L.remove(i)$ \Comment*[r]{{\scriptsize\tt remove itemset at index $i$ that overlaps with the selected itemset $u$.}}
        }{
            $i \leftarrow i + 1$;\\
        }
    }
}
{\bf Return} $S$;
\end{algorithm}

\subsection{Selecting heuristically non-overlapping \MFI}
\label{ii}

To ensure a better compression, we have to select the \MFI for which both the length (or size) and the frequency values are maximized. However, maximizing simultaneously these two conflicting objectives is challenging because the larger the \MFI, the lower its frequency. Instead, we propose to maximize the area criterion since it represents a good compromise between these two criteria. 

Several approaches in the literature use the concept of a tile and its area as an objective interestingness measure for itemsets.
A tile consists of a block of ones in a binary database. For instance, the top$-k$ tiles problem which asks for the $k$ tiles that have the largest area is known to be NP-hard~\cite{tiling} even for $k=1$. 
We propose in this paper a greedy algorithm which finds a sub-set of non-overlapping \MFI maximizing the area.

As pointed out earlier, each tuple of a table constraint can be compressed using at most only one \MFI. To select only non-overlapping \MFI with the largest area values, our algorithm  sorts the \MFI in decreasing order of their area value, selects the first \MFI for compression and removes those for whose coverage overlap with the coverage of the selected \MFI. 
Algorithm~\ref{filter} details the different steps for selecting heuristically the \MFI relevant for compression. 
It exploits a data structure $E$ with two elements, the \MFI $p$ and its area value $area$. A function $Comparator$ is defined to perform a pairwise comparison between itemsets (lines $1$-$10$, Algorithm~\ref{filter}). First, we compute for each \MFI $u$ its area and insert the pair ($u$, $area$) in the list $L$ (lines $12$-$15$, Algorithm~\ref{filter}). Second, we sort the elements of 
$L$ in decreasing order of their area values using the function $Comparator$ (line $16$, Algorithm~\ref{filter}). Finally, we select the first \MFI $u$ from the ordered list $L$, add it to the list $S$ of \MFI relevant for compression, remove $u$ from $L$ and remove from $L$ all the \MFI that overlap with the coverage of $u$ (lines $18-28$, Algorithm~\ref{filter}). This process is repeated until there are no more \MFI to select. 

\begin{table}[t]
\caption{Running \MFIC on Table~\ref{tdc}.}
\begin{minipage}{.5\textwidth}\centering
			\subfloat[{The \MFIs extracted from Table~\ref{tdc} with \smin$ = 2$. Covers and area are also shown.}\label{ex:tab:max}]{
			\scalebox{0.75}{
			\begin{tabular}{llll}	
				\toprule
                ~Idx&~Maximal frequent itemsets & ~Coverage~& Area \\
			  \midrule
			 \textbf{1}&$\langle x_{0}=1, x_{1}=1, x_{2}=2 \rangle$ & $\{t_{8}, t_{9}, t_{10}\}$&9  \\
             \textbf{2}&$\langle x_{0}=0, x_{1}=0, x_{2}=0, x_{4}=2\rangle$ & $\{t_{0}, t_{1}\}$&8 \\
            \textbf{3}&$\langle x_{0}=0, x_{1}=0, x_{2}=1, x_4=2\rangle$ & $\{t_{3}, t_{5}\}$&8\\ 
            \textbf{4}&$\langle x_{0}=0, x_{3}=2, x_{4}=0\rangle$ & $\{t_{2}, t_{4}\}$&6 \\
            \textbf{5}&$\langle x_{0}=1, x_{1}=0, x_{3} = 1, x_{4}=2\rangle$& $\{t_{1}, t_{3}\}$&8\\
            \textbf{6}&$\langle x_{0}=0, x_{2}=1, x_{3} = 3, x_{4}=0\rangle$& $\{t_{7}, t_{10}\}$&8\\
           		\bottomrule
			\end{tabular}
		} 
			}
		\end{minipage}
		\hfill\begin{minipage}{.49\textwidth}
			\centering
			\subfloat[{Sorting the \MFI w.r.t area.}\label{tb:area}]{
			\scalebox{0.75}{
			\begin{tabular}{llll}	
				\toprule
                ~Idx& Area \\
			  \midrule
			   \textbf{1}& $9$\\ 
			 \textbf{2}& $8$ \\
			 \textbf{3}& $8$  \\
             \textbf{5}& $8$ \\
            \textbf{6}& $8$ \\
            \textbf{4}&$6$\\
        	\bottomrule
			\end{tabular} 
		} 
			} 
		\end{minipage}
\end{table}

Sorting the set of \MFI of Table~\ref{ex:tab:max} leads to the ordering of Table~\ref{tb:area}. From this ordering, we select the first \MFI, add it to the set $\mathcal{S}$ and remove from Table~\ref{tb:area} all those that overlap with this \MFI, i.e. the \MFI with $idx=6$. In the second iteration of the while loop of line $17$, the \MFI with $idx=2$ is selected, 
 added to $\mathcal{S} = \{\langle x_{0}=1, x_{1}=1, x_{2}=2\rangle\}$ and the \MFI with $idx=5$ is removed. Finally, when there are no more \MFI to select, Algorithm~\ref{filter} returns the set $\mathcal{S} = \{\langle x_{0}=1, x_{1}=1, x_{2}=2\rangle, \langle x_{0}=0, x_{1}=0, x_{2}=0, x_4=2\rangle, \langle x_{0}=0, x_{1}=0, x_{2}=1, x_4=2\rangle, \langle x_{0}=0, x_{3}=2, x_{4}=0\rangle\}\}$. 
 
 \subsection{Creating the compressed table constraint}
 The last step of \MFIC algorithm consists to create the compressed table constraint $R^c$. This is done by associating an entry for each each \MFI in $\mathcal{S}$. However, as tuples six and seven of Table~\ref{tdc} cannot be compressed using $\mathcal{S}$, a default entry is then created for these two tuples. The final compressed table constraint is showed in Table~\ref{tab:cmp}.

\begin{table}[ht] \centering
\caption{Compressed table constraint. \label{tab:cmp}}
	\begin{center}
\begin{minipage}{.29\textwidth}
		\label{tab:estr}{
			\begin{tabular}{ll} 
              \toprule
                $x_{0}$ $x_{1}$ $x_{2}$ $x_{4}$& \multicolumn{1}{c}{$x_{3}$} \\
               \midrule
              \multirow{2}{*}{0 $\:$ 0 $\:$ 1 $\:$ 2} & 1\\
                &  3 \\
                \bottomrule
                Entry $e_1$.
                \end{tabular}		
		}
		\end{minipage}
		\begin{minipage}{.29\textwidth}
		\centering{
			\begin{tabular}{ll} 
              \toprule
                $x_{0}$ $x_{1}$ $x_{2}$& $x_{3}$ $x_4$ \\
               \midrule
              \multirow{3}{*}{1 $\:$ 1 $\:$ 2} & 0 $\:$ 1\\
                &  2 $\:$ 2 \\
                & 3 $\:$ 0 \\
                \bottomrule
                Entry $e_2$.
                \end{tabular}		
    }
   \end{minipage}
  \begin{minipage}{.29\textwidth}
		\centering{
			\begin{tabular}{ll} 
              \toprule
                $x_{0}$ $x_{1}$ $x_{2}$ $x_4$& $x_{3}$ \\
               \midrule
              \multirow{2}{*}{0 $\:$ 0 $\:$ 0 $\:$ 2} & 0 \\
                &  1 \\
                \bottomrule
                Entry $e_3$.
                \end{tabular}		
    }
   \end{minipage}
    \begin{minipage}{.29\textwidth}
		\centering{
			\begin{tabular}{ll} 
              \toprule
                $x_{0}$ $x_{3}$ $x_{4}$& $x_{1}$ $x_2$ \\
               \midrule
              \multirow{2}{*}{0 $\:$ 0 $\:$ 0 $\:$ 2} & 2 $\:$ 0\\
                &  0 $\:$ 1 \\
                \bottomrule
                Entry $e_4$.
                \end{tabular}		
    }
   \end{minipage}
    \begin{minipage}{.2\textwidth}
    {
			\begin{tabular}{lll} 
           	\toprule
            $x_{0}$ $x_{1}$ $x_{2}$ $x_{3}$ $x_{4}$ \\
            \midrule
            1 $\:$ 0 $\:$ 2 $\:$ 1 $\:$ 2\\
            1 $\:$ 0 $\:$ 2 $\:$ 3 $\:$ 0\\
            \bottomrule
            default table $df$.
        \end{tabular}
     } 
    \end{minipage} 
    	\end{center}
\end{table}

\section{Complexity analysis}

\label{complx}
To analyse the time complexity of \MFIC, we analyse the time complexity of each step. Let $n$ be the number of \MFI, 
\begin{itemize}
   % \item the time complexity to compute the value of \smin with the \topk approach: 
    %
     \item The time complexity of the \topk~\CFIs and \MFI using {\tt LCM} (lines 1, 2 and 3, Algorithm~\ref{filter}) is linear in the number of \CFIs ~\cite{lcm}.
        
%\textcolor{green}{On utilise LCM pour les topk et la complexité de LCM est linéaire}
 %   \item the time complexity to mine \MFI using \lcmmax is\textcolor{orange}{
  %  $O(\sum_{i} (\vert cover (u \cup \{i\})\vert))$ where $u$ a \CFIs,  $i \in \I$ and $i \notin u$ such that $\{u \cup {i}\}$ is infrequent ~\cite{lcm0}}; \textcolor{green}{LCM est d'une complexité lineaire}
   % \item the time complexity to insert $n$ \MFI in a list is $O(n)$;
    \item the sort of line $16$ of Algorithm~\ref{filter}) can be done in $O(n~log~n)$;
    \item  the time complexity to select the \MFI relevant for compression (lines17 to 29, Algorithm~\ref{filter}) is $O(n*((n+1)/2))$ = $O(n^2)$ in the worst case. 
\end{itemize}
So the time complexity of \MFIC is $\theta(n^2)$ where n is the number of \CFIs.

\section{Enforcing GAC on CSP compressed with {\tt MFI-Compression}}
\label{Exploiting}
To enforce GAC on the CSP compressed using the {\tt MFI-Compression} method, we used the \STRST algorithm \cite{b1} wich is an optimized variant of {\tt STR2} that works on compressed table constraints, i.e. a set of entries where each entry consists on an itemset and its corresponding sub-table.\\
To maintain GAC, \STRST checks the validity of entries, where an entry is said to be valid if both of its itemset and at least one tuple of its corresponding sub-table are valid. The method uses a limit pointers to save the index of the latest valid entry and the index of the latest valid sub-tuple of the sub-table corresponding to each valid entry. When restoring entries and sub-tuples, the method just has to modify the value of the limit pointers. \\  We denoted by \STRC the combination of \MFIC with structure of \STRST used to enforce GAC on table constraints.

\begin{examplee}
   Consider the compressed constraint relation of Table~\ref{tab:cmp}. Let \textbf{entriesLimit} (resp. \textbf{limit}) be the index of the last current (valid) entry (resp. the index of the latest valid sub-tuple in the entry). Firstly, all the entries are valid so \textbf{entriesLimit} = 5. \STRST is called after an event is generated. In Table~\ref{tab:solv}, considering that the new event is $x_1$ $\neq$ $0$ (i.e., the removal of the value $0$ from $dom(x_1)$), \STRST starts checking  the validity of the current entries (from 1 to \textbf{entriesLimit}). For the first entry, the itemset is not valid. We do not need to check the validity of its sub-table. We consider the entry as not valid. The second entry is valid because it does not contain $x_1 =0$. Like the entry $e_1$ the entry $e_3$ is not valid. For the entry $e_4$, the validity of the itemset $u = (x_0 = 0,  x_3=2, x_4 = 0)$ is checked. Since $u$ remains valid, the sub-table is scanned. Only the sub-tuple $(x_1 = 0, x_2 = 1)$ remains invalid, thus the value of $limit = 1$. For $df$, the two tuples are not valid. So the value of \textbf{entriesLimit} is $2$. \\
 
\begin{table}[ht] \centering
\caption{{\tt STR-slice} called on a slice table constraint after the event $x_1$ $\neq$ $0$. \label{tab:solv}}
	\begin{center}
\begin{minipage}{.29\textwidth}
		\label{tab:estr}{
			\begin{tabular}{ll} 
              \toprule
                $x_{0}$ $x_{1}$ $x_{2}$ $x_{4}$& \multicolumn{1}{c}{$x_{3}$} \\
               \midrule
             \multirow{2}{*}{\cellcolor{yellow} 0 $\:$ 0 $\:$ 1 $\:$ 2} & 1\\
                &  3 \\
                \bottomrule
                $e_1: limit = 2$.
                \end{tabular}		
		}
		\end{minipage}
		\begin{minipage}{.29\textwidth}
		\centering{
			\begin{tabular}{ll} 
              \toprule
                $x_{0}$ $x_{1}$ $x_{2}$& $x_{3}$ $x_4$ \\
               \midrule
              \multirow{3}{*}{1 $\:$ 1 $\:$ 2} & 0 $\:$ 1\\
                &  2 $\:$ 2 \\
                & 3 $\:$ 0 \\
                \bottomrule
                $e_2: limit = 0$.
                \end{tabular}		
    }
   \end{minipage}
  \begin{minipage}{.29\textwidth}
		\centering{
			\begin{tabular}{ll} 
              \toprule
                $x_{0}$ $x_{1}$ $x_{2}$ $x_4$& $x_{3}$ \\
               \midrule
              \multirow{2}{*}{\cellcolor{yellow} 0 $\:$ 0 $\:$ 0 $\:$ 2} & 0 \\
                &  1 \\
                \bottomrule
                $e_3: limit = 0$.
                \end{tabular}		
    }
   \end{minipage}
    \begin{minipage}{.29\textwidth}
		\centering{
			\begin{tabular}{ll} 
              \toprule
                $x_{0}$ $x_{3}$ $x_{4}$& $x_{1}$ $x_2$ \\
               \midrule
              \multirow{2}{*}{0 $\:$ 2 $\:$ 0} & 2 $\:$ 0\\
                &\cellcolor{yellow}   0 $\:$ 1 \\
                \bottomrule
                $e_4: limit = 1$.
                \end{tabular}		
    }
   \end{minipage}
    \begin{minipage}{.2\textwidth}
    {
			\begin{tabular}{lll} 
           	\toprule
            $x_{0}$ $x_{1}$ $x_{2}$ $x_{3}$ $x_{4}$ \\
            \midrule
          \cellcolor{yellow}   1 $\:$ 0 $\:$ 2 $\:$ 1 $\:$ 2\\
          \cellcolor{yellow}   1 $\:$ 0 $\:$ 2 $\:$ 3 $\:$ 0\\
            \bottomrule
            $df: limit = 0$.
        \end{tabular}
     } 
    \end{minipage} 
    	\end{center}
\end{table}

%\begin{table}[t] \centering
%\caption{{\tt STR-slice} called on a slice table constraint after the event $x_1$ $\neq$ $2$. \label{tab:solv}}
%	\begin{center}
%\begin{minipage}{.33\textwidth}
%		\label{tab:estr}{
	%		\begin{tabular}{lll} 
%              \toprule
%                $x_{0}$ $x_{2}$ $x_{4}$ $x_{5}$& \multicolumn{1}{c}{$x_{1}$ $x_{3}$} \\
%               \midrule
%                & 1 $\:$ 0\\
%                0$\:$ 0 $\:$ 0 $\:$ 0& 1 $\:$ 1 \\
%                & 1 $\:$ 2 \\
%                 & \cellcolor{gray}2 $\:$ 1\\ 
%                \botrule
%                $e_1$: $limit$ = $3$.
%                \end{tabular}		
%		}
%		\end{minipage}\begin{minipage}{.33\textwidth}
%		\centering{
%					\begin{tabular}{lll} 
%						\toprule
 %                   $x_{0}$ $x_{1}$ $x_{2}$ $x_{4}$& \multicolumn{1}{c}{$x_{3}$ 
  %                  $x_{5}$ } \\
 %                   \midrule
%	                & 0 $\:$ 0\\
%                    \cellcolor{gray}0 $\:$ 2 $\:$ 1 $\:$ 0 & 1 $\:$ 1\\ 
%                    & 2 $\:$ 1\\
%                    & 2 $\:$ 0\\
%                 \botrule
%                    $e_2$: $limit$ = $0$.
%            \end{tabular}
%    }
 %  \end{minipage}
  
  %  \begin{minipage}{.33\textwidth}
  %  {
%			\begin{tabular}{lll} 
%           	\toprule
%            $x_{0}$ $x_{1}$ $x_{2}$ $x_{3}$ $x_{4}$ $x_{5}$ \\
%            \midrule
           
%           \rowcolor{gray} 0 $\:$ 2 $\:$ 0 $\:$ 2 $\:$ 0 $\:$ 1\\
%            \botrule
%            $df$: $limit$ = $0$.
%        \end{tabular}
%     } 
 %   \end{minipage} 
  %  	\end{center}
%\end{table}
   \end{examplee}

\section{Experiments}
The experimental evaluation is designed to determine how (in terms of CPU time) \STRC compares to the state-of-the-art of GAC-based algorithms.

\subsection{Experimental protocol}
We performed experiments on the same benchmarks\footnote{Data sets are available at \url{https://bitbucket.org/pschaus/xp-table/src/master/instances/}} used in \cite{b1}. 
Table~\ref{datasets} summarizes the characteristics of each of them. For each benchmark, we give 
the number of its instances ($Ins_{nbr}$), 
the maximum number of variables ($X$) in an instance, 
the largest domain ($|D|$), the largest number of relations ($R_{nbr}$), the size of the largest relation ($R_{max}$), the largest arity of relations ($arity$), the greatest number of constraints ($C_{nbr}$).

\begin{table}[h]
\begin{center}
\begin{minipage}{\textwidth}
\caption{Characteristics of the used benchmarks.}\label{datasets}%
\begin{tabular}{@{}llllllll@{}}
\toprule
	Benchmark& $Inst_{nbr}$ &$X$&$|D|$& $R_{nbr}$ & $R_{max}$&$arity$& $C_{nbr}$ \\
\midrule
  bddLarge& 35 &21 &2 &1 &57971 &18 &133 \\
  bddSmall& 35 & 21 &2 &1 &6945 &15 &2713 \\
  randsJC2500& 10& 40&8 & 40& 2500&7 &40 \\
  randsJC5000& 10& 40& 8&40 &5000 &7 &40 \\
  randsJC7500& 10& 40& 8& 40& 7500& 7&40 \\
  randsJC10000& 10& 40& 8& 40& 10000& 7& 40\\
  Crossword-Lex-Vg& 63& 288& 26&2 & 3607&18 &34 \\
  Crossword-Words-Vg& 65& 320&25 &2 &68064 &20 &36 \\
  Modified-Renault& 50 &111 &42 &142 &48721 &10 &159 \\
\bottomrule
\end{tabular}
\end{minipage}
\end{center}
\end{table}

The implementation of \STRC was carried out in the {\tt Oscar} solver \footnote{Solver available at \url{https://bitbucket.org/oscarlib/oscar/src/dev/}}. The implementation of algorithms of the stat-of-the-art selected for our comparison are also available in the {\tt Oscar} solver. 
All experiments were conducted on Intel (R) Core(TM), $i5-7200$ CPU, $2.5$ GHz with a RAM of $4$ GB, running the Ubuntu $64$ bits $20.04$ LTS operating system. A time limit of $1,800$ seconds has been used per instance. When the runtime exceeds this limit the resolution stops and the instance is considered as failed. 

For our experiments, we fixed the initial value of \smin to 2 for \STRST, contrary to \MFIC that exploits the \topk mining method to fix its value. To be relevant for compression an \MFI must cover at least two tuples and a tuple can be compressed using one and only one \MFI. For this, to fix the value of \smin we varied the value of $k$ on the number of tuples in the table constraint to compress. After several experiments the value of $k$ was varied between 20\% and 60\% of the number of tuples of the table constraint to compress then the value of \smin was set to the average of frequency values returned by the \topk algorithm.
\subsection{Comparing \STRC with \STRST and {\tt STR2}}
The \STRC and \STRST~\cite{b1} are both based on itemsets mining technique for compression and use the same structure (entries) of compressed table constraints. The main differences between the two methods are: (i) \STRC dynamically fix, for each table constraint, the value of the minimum threshold \smin while for \STRST the value of \smin is fixed to 2. (ii) \STRC compresses table constraints using \MFI while \STRST compresses them using frequent itemsets. The two methods are based on {\tt STR-slice}~\cite{b1} to solve the compressed CSP. {\tt STR-slice}~\cite{b1} is an optimized version of {\tt STR2} for compressed CSP. Hence the interest of comparing them.\\
%To mine the \MFI required for compression, the \STRC uses the \topk \CFIs algorithm. The value of $K$ is not known and it has to be fixed. After several experimentation, we varied the value of $k$ between $20\%$ and $60\%$of the number of tuples in the table constraint to compress. This garantie the non 
%%In Table~\ref{tab:STRCST} we report the compression results obtained for all the %benchemarks with \STRC and \STRC. For each benchmark we report the average %compression rate ($c-rate$), the percentage of compressed tuples $c-tup$, the average number of itemsets used for compressing an instance $\vert M \vert$, the average length of itemsets $\vert u \vert$ and the average frequency value of an itemset $\vert freq(u) \vert$.

%\begin{table}[ht]
 %   \centering
  %  \caption{Compression results obtained for the selected benchamarks with the two methods \STRC and \STRST.}\label{cmpp}
  %  \begin{tabular}{@{}ccccccc@{}}
%	\hline\noalign{\smallskip}
 %        benchmark& \multicolumn{3}{c}{\STRC}& \multicolumn{3}{c}{\STRST} \\
  %       \cline{2-7}
   %      &$rate(\%)$&$avg_{fr}$&$avg_{ln}$&$rate(\%)$&$avg_{fr}$&$avg_{ln}$\\
%	\hline\noalign{\smallskip}
 %        bddLarge &13.4 &178 &7 &58 & 246 &3\\
  %       crossword-lexVg &22.1 &9 &4 &29.45 & 3&3\\
   %      crossword-words & 29&14 &4 &18.8&3 &4\\
    %     modifiedRenault &57.8 &20 &3 &37.06  &8 &3\\
     %    randsJC2500 &17.9 &10 & 3&28.4  &3 &3\\
      %   randsJC5000 & 15.6& 17& 3&34.8  &3 &3\\
       %  randsJC7500 &15.6 & 17&3 &38.4 &3 &3\\
        % randsJC10000 &16.1 &29 & 3& 40& 3 &3\\
%    \end{tabular}
%\end{table}

\begin{figure}[ht]
\centering
\includegraphics[width=10cm]{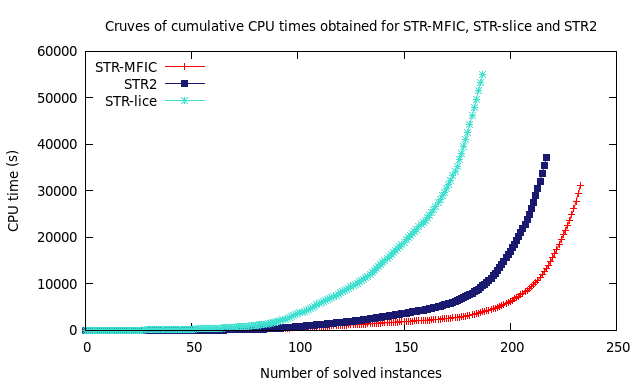}
\caption{The cumulative curves of solving CPU time(s) obtained for \STRC, \STRST and {\tt STRE2} on the selected benchmarks. The x-axis represents the solved instances while the y-axis the cumulative CPU time.}\label{STRSTSTRMFICSTR2}
\end{figure}

In Table~\ref{comp1}, we reported for each method \STRC, \STRST and {\tt STR2} and for each benchmark the number of solved instances ($inst_s$) within 1800s and the average CPU time of solving an instance of each benchmark. 
\begin{itemize}
    \item \textit{number of solved instances}: the three methods solved the same number instances except for the benchmark \textit{randsJC1000} where \STRST did not solved any instance and for \textit{crossword-lexVg} where \STRC and {\tt STR2} solved more instances (7) compared then \STRST. 
    \item \textit{average CPU time}: \STRC performed better compared to \STRST and {\tt STR2} on the average CPU time required to solve each instance of the different benchmarks except for \textit{crossword-lexVg} where the average CPU time required by \STRST is less then the one required by \STRC and {\tt STR2}. Also for \textit{randsJC5000} and \textit{randsJC7500}, {\tt STR2} solved each instance on average CPU time less than that of \STRC and \STRST.
\end{itemize}
Figure~\ref{STRSTSTRMFICSTR2} shows the cumulative curves of the solving CPU time(s) obtained for \STRC, \STRST and {\tt STR2} for the selected benchmarks. We remark that for the 60 first instances, the three cumulative curves are identical, then the curves of {\tt STR2} and \STRC dominate that of \STRST. The two cumulative curves of {\tt STR2} and \STRC are very close for the 150 first instances then the curve of \STRC dominates that of {\tt STR2}. \STRC solved more instances compared to \STRST and {\tt STR2}.   
\begin{table}[ht]
    \centering
    \caption{Comparing the \STRC, \STRST and {\tt STR2} on the number of solved instances ($inst_s$) and average solving CPU time ($CPU_t$)}\label{comp1}
    \begin{tabular}{@{}ccccccc@{}}
    \toprule
        benchmark& \multicolumn{2}{c}{\STRC} & \multicolumn{2}{c}{\STRST} & \multicolumn{2}{c}{{\tt STR2}} \\
       \cline{2-7}
        &$inst_s$&$CPU_t(s)$&$inst_s$&$CPU_t(s)$&$inst_s$&$CPU_t(s)$\\
   \midrule
       bddLarge&\textit{35}&\textbf{60}&\textit{35}&382&\textit{35}&65\\
      bddSmall&\textit{35}&\textbf{28}&\textit{35}&195&\textit{35}&37\\
     crossword-lexVg&\textit{35}&338&\textit{28}&\textbf{325}&\textit{35}&352\\
     crossword-words&\textit{23}&\textbf{33}&\textit{23}&157&\textit{23}&46\\
     modifiedRenault&\textit{39}&\textbf{72}&\textit{39}&88&\textit{39}&150\\
       randsJC2500&\textit{10}&\textbf{11}&\textit{10}&35&\textit{10}&12\\
        randsJC5000&\textit{10}&154&\textit{10}&553&\textit{10}&\textbf{130}\\
       randsJC7500&\textit{10}&636&\textit{10}&1373&\textit{10}&\textbf{563}\\
        randsJC10000&\textit{10}&\textbf{703}&\textit{0}&TO&\textit{10}&750\\
    \bottomrule
    \end{tabular}
\end{table}
Even if both \STRC and \STRST compress the table constraints before there solving, \STRC behaved better on most benchmarks. To clarify the obtained results, Table~\ref{tab:STRCST} gives some details about the compression process of some selected benchmarks. For each benchmark and for each method, we reported, in percentage ($\%$), the number of compressed tuples (\textit{c-tup}), the compression rate (\textit{c-rate}), the average number of itemsets (\textit{$\vert M\vert$}), frequent itemsets for \STRST and \MFIs for \STRC, mined from each instance of a benchmark, the average length (\textit{$\vert u \vert$}) and the average frequeny value (\textit{$freq(u)$}) of each itemset $u$. \STRST compressed more tuples and offered better compression rate compared to \STRC. But when comparing the number and the frequency of itemsets used for compression, we can see that \STRST used a very large number of itemsets with low frequencies while \STRC used less number of itemsets with high frequencies. For example for the benchmark \textit{randsJC2500}, \STRC compressed about 20\% of tuples of each instance with only 86 \MFIs with an average frequency equals to 16. \STRST compressed about 20\% more tuples compared to \STRC, using 470 frequent itemsets with an average frequency equals to 3. So, \STRST compress table constraint with a very large number of frequent itemsets with smallest frequencies, therefore the resulting compressed table is composed of a large number of smallest entries. Solving compressed table constraints with high number of smallest entries can slow down the solving process due to the number of entries to iterate.\\
\begin{table}[ht]
    \centering
    \caption{Comparing \STRC and \STRST on number of compressed tuples (\textit{c-tup}), compression rate (\textit{c-rate}), number of itemsets ($\vert M\vert$), length of itemsets ($\vert u \vert$) and frequency of itemsets ($freq(u)$).} \label{tab:STRCST}
    \begin{tabular}{@{}ccccccc@{}}
   \toprule
    benchmark& method &c-{tup}(\%)&c-rate(\%)&$\vert M\vert$&$\vert u \vert$&$\vert freq(u)\vert$ \\
    \midrule
     \multirow{2}{2cm}{Crossword-LexVg}& \STRC & 44.3 & 22.1 & 88 & 4 & 18 \\
                                      &\STRST &57.13 & 29.45 & 667 & 3 & 2 \\ \\
    \multirow{2}{2cm}{Crossword-WordsVg}&\STRC& 52.04 & 29.04 & 95 & 4 & 30 \\
                                        & \STRST & 69.4 & 18.8 & 376 & 4 & 3\\\\
     \multirow{2}{2cm}{randsJC2500}& \STRC &39.2 & 17.9 & 86 & 3 & 16 \\
                                   & \STRST & 59.27 & 28.04 & 470 & 3 & 3\\\\
     \multirow{2}{2cm}{randsJC5000}& \STRC & 38.14 & 15.6 & 98 & 3 & 24 \\
                                  & \STRST &71.26 & 34.87 & 1039 & 3 & 3 \\\\
    \multirow{2}{2cm}{randsJC7500}& \STRC &  36.17 & 15.8 & 91.8 & 3 & 28 \\
                                  & \STRST & 76.44 & 38.4 & 1604 & 3 & 3 \\ \\
    \multirow{2}{2cm}{randsJC10000} & \STRC & 35.5 & 16.1 & 91 & 3 & 32 \\
                                    &\STRST & 79.54 & 40.91 & 2177 & 3 & 3 \\\\
    \multirow{2}{2cm}{bddLarge}& \STRC & 24.5 & 13.4 & 25 & 7 & 82 \\
                               &\STRST &91&58&1655&6&3 \\
    \bottomrule
    \end{tabular}
\end{table}
%the benchmarks The compression rate obtained with \STRST is higher than that obtained with \STRC and \STRST compressed more tuples compared to \STRC, but we can see that \STRST used a very large number of frequent itemsets with very small frequency values while \STRC compressed the same benchmark with less number of \MFIs that have 

%%%%%%%%%%%%%%%%%%%%%%%%%%%%%%%%%%%%%%%%%%%%%%%%%%%%%%%%%%%%%%%%%%%%%%%%
%
\subsection{Comparing \STRC with state-of-the-art {\tt GAC}-based algorithms}
Our last experiment aims at comparing our approache \STRC with state-of-the-art algorithms enforcing GAC on table constraints. The tested GAC algorithms are {\tt STR3}~\cite{LecoutreLY15}, {\tt shortSTR2}~\cite{JeffersonN13}, {\tt STRBit}~\cite{WangXYL16}, {\tt MDD4R}~\cite{PerezR14}, {\tt GAC4}~\cite{MohrM88}, {\tt GAC4R}~\cite{GAC4R} and {\tt CT}~\cite{VerhaegheLS17}. 
Figure~\ref{gnuplotall} depicts the curves of cumulative CPU times obtained for the \STRC method (the red curve) and the selected GAC algorithms of the state-of-the-art among all instances of the used benchmarks. Clearly, compression approaches based on Bitwise representation such as {\tt CT} and {\tt STRBit} are the best performer methods with a slight advantage to {\tt CT}. Even if {\tt STRbit} dominates our \STRC method, we notice that their corresponding curves get closer and closer until they are almost identical after 230 solved instances. the {\tt shortSTR2} and \STRC are competitive such that their cumulative curves are almost identical for the first 180 solved instances then we can clearly see that the cumulative curve of \STRC dominates that of {\tt shortSTR2} and solved more instances. Comparing to the other selected {\tt GAC}-based algorithms, the cumulative curve of \STRC dominates all the others cumulative curves and solves more instances.    

 \begin{figure}[ht]
\centering
\begin{minipage}[b]{0.8\textwidth}
		\label{mod0}
		\includegraphics[width=\textwidth]{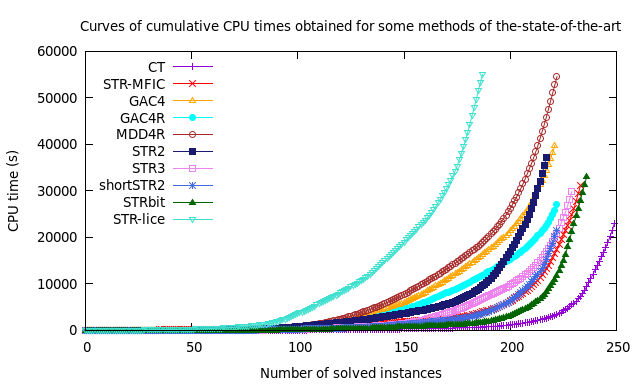}
		\caption{The cumulated CPU time (s) of selected GAC algorithms. The x-axis represents the solved instances and the y-axis the total solving time. There are 250 instances from 9 divers benchmarks}\label{gnuplotall}
	\end{minipage}
	\begin{minipage}[b]{0.8\textwidth}
		\label{mod1}
		\includegraphics[width=\textwidth]{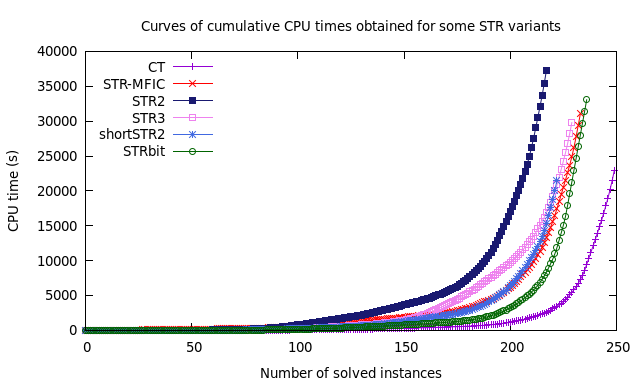}
		\caption{The cumulated CPU time (s) of our \STRC and some STR variants algorithms.}\label{gnuplotall}
	\end{minipage}
	
%\includegraphics[width=10cm]{figures/all.png}
%\caption{The cumulated CPU time (s) of selected GAC algorithms. The x-axis represents the solved instances and the y-axis the total solving time. There are 250 instances from 9 divers benchmarks}\label{gnuplotall}
\end{figure}

\begin{figure}[ht]
\centering
\begin{minipage}[b]{0.48\textwidth}
		\label{mod0}
		\includegraphics[width=\textwidth]{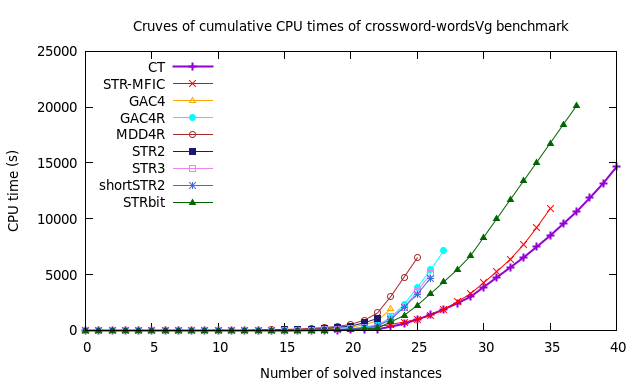}
	\end{minipage}
	\begin{minipage}[b]{0.48\textwidth}
		\label{mod1}
		\includegraphics[width=\textwidth]{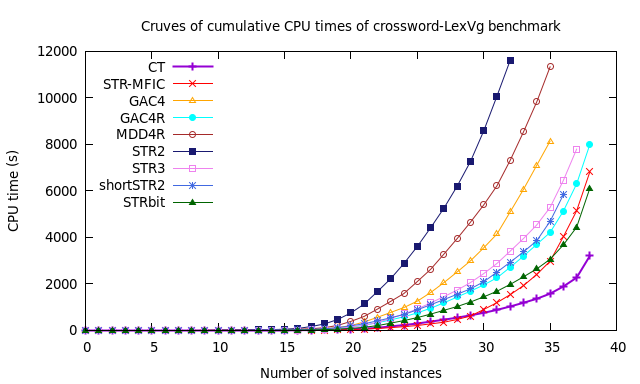}
	\end{minipage}
	\begin{minipage}[b]{0.48\textwidth}
		\label{mod3}
		\includegraphics[width=\textwidth]{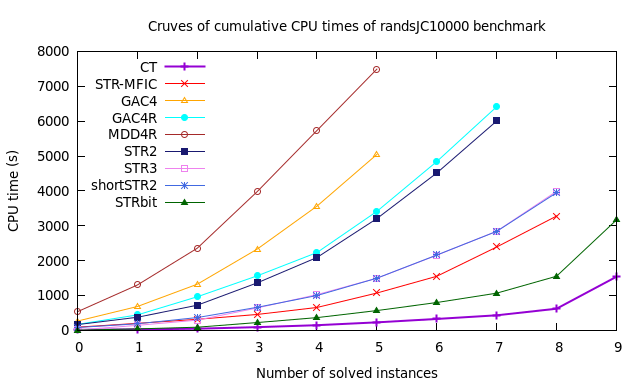}
	\end{minipage}
\caption{Curves of cumulative CPU times obtained for \STRC and the state-of-the-art {\tt GAC}-based methods for the benchmarks \textit{crosswords-words-vg}, \textit{crosswords-lex-vg} and \textit{randJC10000}.}\label{mod}
\end{figure}

In Figure~\ref{mod} depicts the curves of cumulative CPU times obtained for three benchmarks chosen arbitrarily from those selected to conduct our experiments. \\
We can see that for the three benchmarks, the curves of cumulative CPU times of \STRC~\cite{b1} dominate that of \STRST, {\tt STR3}~\cite{LecoutreLY15}, {\tt shortSTR2}~\cite{JeffersonN13}, {\tt MDD4R}~\cite{PerezR14}, {\tt GAC4}~\cite{MohrM88} and {\tt GAC4R}~\cite{GAC4R} and solve more instances.\\
While Comparing to
\begin{itemize}
    \item {\tt STRbit}~\cite{WangXYL16}: for the benchmark \textit{Crossword-words-vg}, even if {\tt STRbit}~\cite{WangXYL16} solved more instances, the curve of cumulative CPU times of \STRC dominates that of {\tt STRbit}~\cite{WangXYL16}. For the benchmark \textit{Crossword-lex-vg}, the two methods \STRC and {\tt STRbit}~\cite{WangXYL16} solve the same number of instances and their cumulative curves are competitive. For the 35 first instances, the cumulative curve of {\tt STRbit}~\cite{WangXYL16} dominates that of \STRC then the one of \STRC dominates it. Finaly, even if {\tt STRbit}~\cite{WangXYL16} and \STRC solved the same number of instance of the benchmark \textit{randsJC10000}, the curve of cumulative CPU times of {\tt STRbit}~\cite{WangXYL16} dominates that of \STRC.
    \item {\tt CT}~\cite{VerhaegheLS17}:  for the benchmarks \textit{Crossword-words-vg} and \textit{Crossword-lex-vg}, the two curves of cumulative CPU times of \STRC and {\tt CT}~\cite{VerhaegheLS17} are almost identical for the first 28 instances of each benchmark, then we remark that the ones of {\tt CT}~\cite{VerhaegheLS17} dominate that of \STRC.
\end{itemize}
%In Figure ~\ref{mod} we give for each of the datasets \textit{crossword-words-vg}, \textit{crossword-lex-vg} and \textit{randsJC1000} the cumulative curves obtained for our {\tt MFI-Compression} method the selected GAC methods. \\

%For the \textit{crossword-lex-vg} dataset, the two cumulatives curves corresponding to {\tt MFI-Compression} and {\tt Compact-table} methods are identical for the 30 first instances then the one of {\tt Compact-table} dominates that of {\tt MFI-Compression}. While the cumulative curve of {\tt MFI-Compression} dominates the one of {\tt STR-bit} for the 35 instances then the cumulative of {\tt STR-bit} starts to slightly dominate that of {\tt MFI-Compression}.

%The second benchmark is the \textit{crossword-words-vg}, one can see that the cumulative curve of the \STRC method and that of {\tt CT} are identical for the 25 first instances then the curve of {\tt Compact-table} dominates that of \STRC.\\

%The third one is the \textit{randsJC10000}, only the cumulative curves of {\tt STRbit} and {\tt Compact-table} dominate that of \STRC.\\

\section{Conclusion}
\label{conc}
In this paper, we have proposed a new approache based on data mining techniques for compressing table constraints. Our approach, called \MFIC enumerates from a table constraint the maximal frequent itemsets (\MFI) relevant for compression. To cope with the problem of fixing the minimum support \smin, we proposed to use the \topk approach. The coverages of \MFI in a table constraint can overlap each other, a tuple of a table constraint can be compressed using one and only one \MFI. To respect this condition and compress a table constraint more effeciently, we proposed to select from the set of \MFI only those having the largest area and do not overlap each other. To solve the compressed CSP, we used the {\tt STR-slice}\cite{b1} which is an optimized variant of {\tt STR2} for compressed CSP. We called the combination of the two methods {\tt MFI-Compression} with {\tt STR-slice} by \STRC.
We evaluated our contributions on different benchmarks, and compared it to some GAG-based methods of the state-of-the-art, namely the different variant of STR ({\tt STR2}, {\tt STR3}, {\tt STR-Slice}, {\tt shortSTR2} and {\tt STRbit}), {\tt GAC4}, {\tt GAC4R}, {\tt MDD4R} and {\tt Compact-table}. The results showed that compressing table constraints using \STRC enables to solve the resulting CSP in less time compared to \STRST \cite{b1} also based on a data mining technique. The results obtained for our method are competitive with that obtained for the other selected GAC-based methods of the state-of-the-art except for {\tt CT}~\cite{VerhaegheLS17} which obtains better results in most cases. As future work we will try find an efficient solution to fix the value \smin.

%\bmhead{Acknowledgments}
%This work has been sponsored by the General Directorate for Scientific Research and Technological Development, Ministry of Higher Education and Scientific Research (DGRSDT), Algeria)

\bibliographystyle{unsrt}  


\end{document}